\documentclass[11pt,preprint]{aastex}

\usepackage{graphics}
\usepackage{lineno}
\usepackage{color}

\shorttitle{Inflow in SDSS J1228+1005}
\shortauthors{Li \textit{et al.}}

\begin{document}

\title{Feeding the Accretion Disk from the Dusty Torus in a Reddened Quasar}

\author{Ge Li\altaffilmark{1}, Xiheng Shi\altaffilmark{2}, Qiguo Tian\altaffilmark{2}, Luming Sun\altaffilmark{3}, Xinwen Shu\altaffilmark{3}, Xiangjun Chen\altaffilmark{1} and Hongyan Zhou\altaffilmark{4,2}}

\altaffiltext{1}{Hefei National Laboratory for Physical Sciences at Microscale and Department of Modern Physics, University of Science and Technology of China, Hefei, Anhui 230026, China}
\altaffiltext{2}{Polar Research Institute of China, 451 Jinqiao Road, Shanghai, China}
\altaffiltext{3}{School of Physics and Electronic Information, Anhui Normal University, Wuhu, Anhui, China}
\altaffiltext{4}{Department of Astronomy, University of Science and Technology of China, Hefei, Anhui 230026, China, mtzhou@ustc.edu.cn}

\begin{abstract}
  We present here a detailed analysis of an unusual absorption line system in the quasar SDSS J122826.79+100532.2.
  The absorption lines in the system have a common redshifted velocity structure starting from $v\sim0$ and extending to $\sim1,000\ \mathrm{km~s}^{-1}$, and are clearly detected in hydrogen Balmer series up to H$\iota$, in metastable neutral helium triplet, and in optical lines of excited states of single ionized iron.
  We estimated that the absorber has a density $n_{\mathrm{H}}\approx10^{8.4}\ \mathrm{cm}^{-3}$ and an ionization parameter $U\approx10^{-1.2}$, thereupon located it at $r_{\mathrm{abs}}\approx1.5$ pc from the central supermassive black hole.
  The inferred distance is remarkably similar to the evaporation radius for dust grains $r_{\mathrm{evap}}\approx1$ pc in the quasar.
  Thus the absorber may be a probe of an inflow starting from the dusty torus and feeding the accretion disk.
  Both the featureless continuum and the broad emission lines are heavily reddened with $E(B-V)\approx0.66$, in contrast to the narrow emission lines whose reddening is negligible.
  The dusty medium could be located in between the broad and narrow emission line regions, and possibly be associated with a 'cold' narrow absorption line system detected in \ion{Ca}{2} and \ion{Na}{1} doublets nearly unshifted from the quasar systemic velocity.
  SDSS J122826.79+100532.2 might represent such a rare case that both the inflow and the torus could be tracked by absorption lines.
\end{abstract}



\section{Introduction}

Super-massive black holes (SMBHs) are found to be ubiquitous in massive galaxies as clearly revealed in active galaxies (AGNs), and as revealed by the motion of gas and stars within the SMBH sphere of influence in normal galaxies. SMBHs are believed to play a key role in the evolution of galaxies. They grow rapidly through merge and accretion of inter-stellar medium (ISM) during the epochs when their host galaxies grow rapidly. The accretion of ISM can release huge amount of energy, activating the galaxies, pouring radiation and feedback, i.e. outflow and jet. It is suggested that the feedback would further affect the star formation in the host galaxies to regulate the properties of galaxies that we observe in the local universe, which explains the tight correlations between the properties of the galactic spheroids and those of their central SMBHs (Kormendy \& Richstone, 1995; Magorrian et al., 1998; Gebhardt et al. 2000).

Though being a fundamental process, the accretion of ISM onto SMBHs remains far from clear. ISM falling inward has been detected at the galactic scale ($\ge 100~\mathrm{pc}$) by Integral Field Unit (IFU) using emission lines (Dumas et al., 2007; Stoklasov\'{a} et al., 2009; Rodr\'{i}guez-Zaur\'{i}n et al., 2011; Storchi-Bergmann \& Schnorr-M\"{u}ller 2019). However, except for a couple of nearby targets (e.g., Fathi et al., 2006; Dumas et al., 2007; Storchi-Bergmann et al., 2007; Riffel et al., 2008; Davies et al., 2009; Schnorr M\"{u}ller et al., 2011), this method is unavailable for the vicinity of SMBHs ($< 10~\mathrm{pc}$) as those regions are now spatially unresolvable in most luminous AGNs.

The emission lines and the absorption lines in spectra present robust probes to the gaseous medium at this scale. Particularly, the redshifted absorption lines with respect to the systemic redshifts of quasars provide a secure probe to the infalling gas. Some cases have been identified (e.g., Hall et al. 2002; Hall et al. 2013) since large sky area surveys, such as the Sloan Digital Sky Survey (SDSS, York et al. 2000) and the Large Sky Area Multi-Object Fiber Spectroscopic Telescope (LAMOST, Cui et al. 2012), became available. Hall et al. (2013) presented a sample of 17 redshifted \ion{C}{4} and \ion{Si}{4} broad absorption line (BAL) quasars, though with the high-ionization absorption lines it is difficult to distinguish between fast inflows and rotationally dominated outflows. A major progress was made recently by Zhou et al. (2019), which reported the discovery of the redshifted BALs and mini-BALs in the optical absorption lines of \ion{H}{1} Balmer (\ion{H}{1}*) and metastable \ion{He}{1} (\ion{He}{1}*) in a sample of eight quasars. With \ion{H}{1}* and \ion{He}{1}* as powerful trackers, the inflows were confirmed in at least two cases: in J112526.12+002901.3 the absorbing gas is falling inward near the inner surface of the dusty torus (Shi et al. 2016a); while in J103516.20+142200.6 Zhou et al. (2019) claimed the first discovery of inflow directly fueling the SMBH, as the absorbing gas is near the edge of the accretion disk.
ssss
The redshifted BALs, both of high-ionization and low-ionization, are far rarer than the blueshifted BALs which represent outflows. This might be a selection effect due to the obscuration, possibly by the dusty torus presumed by the AGN unification schemes (e.g., Antonucci 1993; Ramos Almeida \& Ricci 2017). Since inflows are suggested to lie close to the mid-plane of the central engine, they would be easier to be obscured by the torus than outflows. Therefore, a correlation between the redshifted absorptions and the obscuration (extinction and reddenning) would be expected. However, in the cases of J112526.12+002901.3 and J103516.20+142200.6, both objects are not severely reddened ($E(B-V)\le 0.15$) compared to the SDSS quasar composite spectrum (Vanden Berk et al. 2001). The explanation is that the lines of sight toward the two objects occasionally pass through the holes in the edge of their clumpy dusty tori where the number densities of the dusty clouds are low.

SDSS J122826.79+100532.2 (hereafter as J1228+1005) is one of the eight redshifted \ion{H}{1}* and \ion{He}{1}* BAL or mini-BAL quasars identified in the SDSS quasar catalogue (see Zhou et al. 2019). The absorption lines in the rest-frame optical band show similar pattern with the absorptions in J112526.12+002901.3 and J103516.20+142200.6. However, the spectral energy distribution (SED) is remarkably different. The continuum in the rest-frame optical band is reddened, indicating severe extinction. Therefore, this object may present an example more typical for quasars that host inflows producing redshifted absorption lines.
In \S\ref{Observation} we describe the data of SDSS and P200 TripleSpec spectroscopic observations and the reduction of the latter. In \S\ref{SED} we fit the unusual SED and interpret the results. In \S\ref{Absorptions} we measure the redshifted and blueshifted absorption lines in the SDSS spectrum and estimate the physical properties of the redshifted absorber using photo-ionization simulations. In \S\ref{Discussion} we estimate the properties of the central SMBH, argue whether the inflows will be sufficient to power the central engine or not if the redshifted absorber is a part of them, discuss the possible variation of the redshifted absorptions, discuss the association of the blueshifted absorber with the dusty medium, and give a overall picture of the absorbers in J1228+1005. A brief summary is presented in \S\ref{Summary}. Throughout this paper we assume a cosmology with $H_{0}=70\mathrm{km~s}^{-1}\mathrm{Mpc}^{-1}$, $\Omega_{\mathrm{M}}=0.3$ and $\Omega_{\mathrm{\Lambda}}=0.7$.

\section{Observation and Data Reduction}\label{Observation}

The SDSS spectrum for J1228+1005 was obtain on 2012-01-23 by the SDSS-III Baryon Oscillation Spectroscopic Survey (BOSS; Dawson et al. 2013), covering 3589 to 10354 \AA  in the observer's frame with a mean signal-to-noise ratio of $S/N \sim 13$. The spectroscopic flux is slighly different from the SDSS photometry flux, larger by $\sim 2\%$ at the SDSS $g$-band while by $\sim 17\%$ at the $z$-band. Since the Catalina survey{\footnote{http://nesssi.cacr.caltech.edu/DataRelease/}} shows no long-term variability of the object in the $V$-band from 2005 to 2013, we recalibrated the spectroscopic flux to fit the point-spread function (PSF) magnitudes, by multiplying a quadratic polynomial $a\lambda^2+b\lambda+c$ where $a$, $b$ and $c$ are parameters and $\lambda$ is the observed wavelength in \AA.

A near-infrared (NIR) spectroscopic observation for J1228+1005 was performed on 2017-02-09, using the TripleSpec spectrograph mounted on the Hale 200-inch telescope. Six exposures, each of 180 seconds, were taken, reaching an overall S/N of 4.7. The width of the slit was 1 \arcsec. The data was reduced with the specX package. The wavelength coverage is from 0.97 to $2.46~\mu\mathrm{m}$ in the observer's frame, with two sections around 1.35 and $1.85~\mu\mathrm{m}$ being excluded owing to strong telluric absorptions. The fluxes between 0.97 and $1.35~\mu\mathrm{m}$ was calibrated according to the UKIDSS $Y$ and $J1$ bands' aperture flux 4 photometry by multiplying a linear polynomial $a'\lambda+b'$ (Hodgkin et al. 2009).

The spectra were corrected for the Galactic extinction using the dust maps by Schlegel et al. (1998) and the mean extinction curve of the Milky Way by Fitzpatrick \& Massa (2007).

J1228+1005's SDSS spectrum shows very strong narrow emission lines of [\ion{O}{2}] $\lambda 3728$ and [\ion{O}{3}] $\lambda\lambda 4960,5008$ doublets, compared to the quasar composite (Vanden Berk et al. 2001). Since the [\ion{O}{3}] emission peaks show more complicated profile, containing multiple components which might be associated with outflow, and are superposed on the red wing of the broad H$\beta$ emission line, we chose the [\ion{O}{2}] peak to determine the the systemic redshift. The [\ion{O}{2}] emission in fact consists of a doublet centered at 3727.09 and $3729.88~\mathrm{\AA}$ respectively, with the flux ratio varying normally between 0.8:1 and 0.9:1 in quasars. Therefore, the peak was fitted using a single Gaussian with an effective rest wavelength of $3728.60~\mathrm{\AA}$ (Hewett \& Wild, 2010). The resultant systemic redshift is $0.66843\pm 0.00004$, where the uncertainty of the effective wavelength is not included. The final resultant spectra in the rest-frame are plotted in Figure \ref{SED_fig}, along with the photometry data obtained from GALEX, SDSS, and UKIDSS.

\section{Broad Band Spectral Energy Distribution}\label{SED}

Compared to the quasar composite spectrum (Vanden Berk et al. 2001), the SDSS spectrum is quite red with its flux smaller at the blue end than at the red end, indicating that J1228+1005 is considerably reddened. However, the flux does not continue to decline but goes up shortward rest-frame $3000~\mathrm{\AA}$, which is inconsistent with usual extinction law of always more extinction at shorter wavelengths. This is not a flux calibration issue, as the SDSS $u$ band photometry demonstrates. Also the GALEX NUV photometry shows even larger flux at  $\sim 1360~\mathrm{\AA}$ in the rest-frame. Therefore, the object presents a `V'-shaped broad band spectral energy distribution (SED).

Such an SED reminds us of some AGNs studied previously which have similar SEDs, such as PKS 2355-535 (Scarpa \& Falomo 1997), OI 287 (Li et al. 2015), SDSS J000610.67+121501.2 (Zhang et al. 2017b), and SDSS J120300.19+162443.7 (Pan et al. 2019).
Zhang et al. (2017b) explained that the `V'-shaped SED can be a combination of two components: a `red' one and a `blue' one. The `red' component dominates the radiation at longer wavelengths, and it is contributed by the reddened radiation of the AGN. The `blue' component dominates the excess radiation at shorter wavelengths, but its origin is different with that of the `red' one. Since the on-sky diameter of the fibres feeding the BOSS spectrograph, which is 2 \arcsec,  corresponds to a linear scale of $7.0\times 10^3~\mathrm{pc}$ at the redshift of the object, there could be polluting starlight from the host galaxy posing as the `blue' component. However, star-forming galaxies which present strong radiation at rest-frame $\sim 1360~\mathrm{\AA}$ would also show obvious optical-NIR bumps longward $\sim 4000~\mathrm{\AA}$ (Silva et al. 1998, Polletta et al. 2007, Zhang et al. 2017a). Since such bump can not be identified in J1228+1005, the possibility that the `blue' component comes from the starlight of the host galaxy can be excluded.

\begin{figure}
\includegraphics[width=\columnwidth]{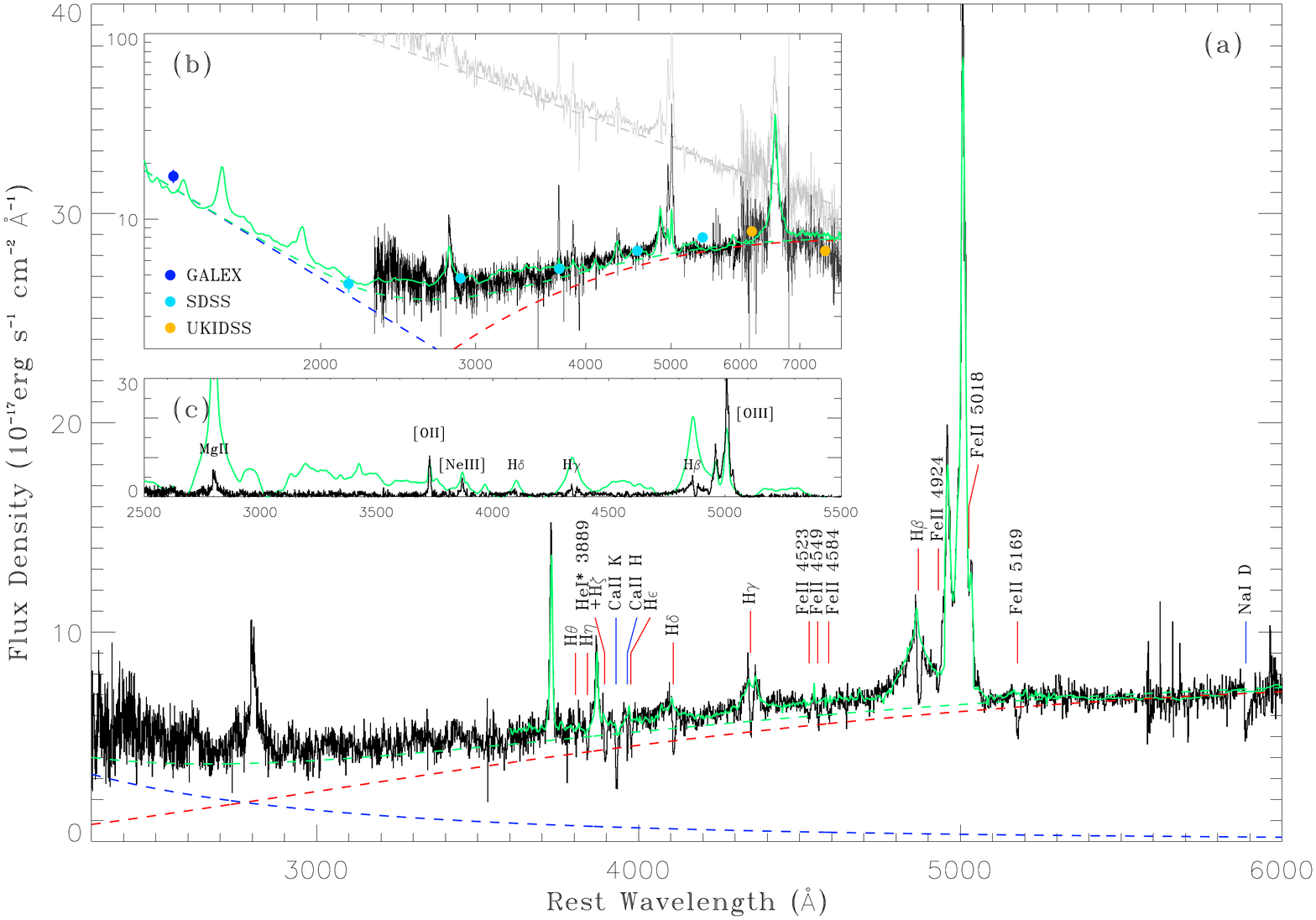}
\caption{Panel (a): The black line presents the BOSS spectrum for J1228+1005. The transitions in the redshifted and blueshifted absorption line systems are identified. The solid green line is the unabsorbed template constructed using the `pair-matching' method. Panel (b): The SED fit for J1228+1005. The black line presents the BOSS spectrum. The dark grey line presents the TripleSpec spectrum. And the solid circular data points present the photometric fluxes from GALEX, SDSS and UKIDSS. The solid green line show the best fit employing the SDSS quasar composite, assuming a reddened component due to dust extinction and a blue component due to scattering. The dashed red and blue lines present the continuum corresponding to the red and blue components, and the dashed green line is the sum. The light grey line show the intrinsic flux of the object with the extinction corrected and the scattering flux removed. Panel (c): The emission lines extracted from the BOSS spectrum, compared with the emission from the composite. The narrow [\ion{O}{2}] and [\ion{O}{3}] lines have fluxes approximately equal to those from the composite, while the broad (\ion{H}{1} Balmer and \ion{Mg}{2}) emissions are much weaker.\label{SED_fig}}
\end{figure}

For SDSS J000610.67+121501.2, Zhang et al. (2017b) suggested that the dominating component at short wavelengths is the central radiation scattered by the ambient medium to where the lines of sight are different from that directly toward the central engine and therefore is not obscured by the dusty medium. In a couple of sources with similar SEDs, such as the SEDs of PKS 2355-535 (Scarpa \& Falomo 1997), O I 287 (Li et al. 2015), and SDSS J091501.71+241812.1 (Yang et al. in preparation), polarized radiations have been confirmed by the spectropolarimetry observations, which supports the scattered radiation paradigm. Therefore, Zhang et al. (2017b) used a model consisting of a reddened composite and a scattered composite to reproduce the rest-frame UV and optical spectra and the photometry for J000610.67+121501.2. This picture also seems reasonable for J1228+1005, so following Zhang et al. (2017b), we used the same formula to fit our spectra and photometry:
\begin{equation}
F_{\lambda}=C_1 (\lambda/3000\ \AA)^{-\alpha_{\mathrm{scat}}}F_{\mathrm{composite},\lambda}+C_2 A(E(B-V),\lambda)F_{\mathrm{composite},\lambda} ,
\end{equation}
where $C_1$ and $C_2$ are the scale factors for the components, $(\lambda/3000\ \AA)^{-\alpha_{\mathrm{scat}}}$ represents the scattering law, $A(E(B-V),\lambda)$ is the extinction law, and $F_{\mathrm{composite},\lambda}$ is the quasar composite of Vanden Berk et al. (2001).
Employing the SMC-type extinction law (Gordon et al. 2003) and masking the emission lines, we find $\alpha_{\mathrm{scat}}=-1.2$, $E(B-V)=0.66$, and $C_1/C_2=0.027$.
The result is present in Figure \ref{SED_fig} panel (b). The model can reproduce the NUV, optical and NIR spectral and photometric fluxes.

The emission lines were not included in the SED fit. However, the Balmer peaks which are dominated by broad emission lines, such as H$\alpha$, H$\beta$, and H$\gamma$, are consistent with those in the SED model. Meanwhile, the narrow emission lines of [\ion{O}{2}] $\lambda 3728$, [\ion{Ne}{3}] $\lambda 3869$, [\ion{O}{3}] $\lambda\lambda 4960,5008$ are much stronger than those in the SED model. In Figure \ref{SED_fig} panel (c), we make further comparison between the observed emission flux (with the reddened and the scattered continuum being subtracted) and the emission flux in the quasar composite (scaled to match the object's intrinsic spectrum). Although the Balmer emission lines are very weak, [\ion{O}{2}] $\lambda 3728$ and [\ion{O}{3}] $\lambda\lambda 4960,5008$ show fluxes roughly equal to those of the scaled quasar composite. This supports that the dusty medium obscures the accretion disk and the broad emission line region (BELR), but not the narrow emission line region (NELR).

In principle, either electrons or dust grains (or them both) may serve as the scattering mirror.
The scattered light shortward of $\sim3000$ \AA\ amounts to about 3\% of the quasar's intrinsic flux at these wavelengths.
If such a strong scattered light were caused mainly by electrons, the mirror would have a column density $N_H\gtrsim 10^{23}\ \rm{cm}^{-2}$
with a global covering fraction $C_f\gtrsim 10\%$.
Much less extreme conditions would, instead, be required for scattering dust.
Furthermore, the scattered light is relatively blue as indicated by the best-fit $\alpha_{\mathrm{scat}}=-1.2$, suggesting that it is not reddened by dust.
We will further discuss in Section 5.4 the obscuring dust and the scattering dust, and the possible relation between them.

\section{The Absorption Line Systems}\label{Absorptions}

The object caught our attention for the detection of the redshifted mini-BALs for \ion{H}{1}* and \ion{He}{1}* in its SDSS spectrum. The absorption lines extend from 0 to $1200~\mathrm{km~s}^{-1}$ in the quasar's rest frame, consisting of obvious troughs for H$\beta$, H$\gamma$, H$\delta$, H$\epsilon$, H$\zeta$, H$\eta$, H$\theta$, H$\iota$, and \ion{He}{1}* $\lambda 3889$. There are also absorption troughs at rest-frame 4531.7, 4558.6, 4593.0, 4933.8, 5028.5, and $5179.4~\mathrm{\AA}$. Comparing them with the stellar wind absorptions in the emission-line star Hen 3-209 (Naz\'{e} et al. 2006) and the \ion{H}{1}*/\ion{He}{1}*/\ion{Fe}{2} absorption system in SDSS J125942.80+121312.6 (Shi et al. 2016b), we identified that these absorptions come from the excited levels of Fe$+$: \ion{Fe}{2}* $\lambda 4523$ from b$^4$F$_{5/2}$, \ion{Fe}{2}* $\lambda 4549$ from b$^4$F$_{7/2}$, \ion{Fe}{2}* $\lambda 4584$ from b$^4$F$_{9/2}$, and \ion{Fe}{2}* $\lambda,\lambda 4924,5018,5169$ from a$^6$S$_{5/2}$, of the same redshifted absorption line system.

Although $\lambda_{\mathrm{rest}} f$ of H$\beta$ is larger than that of H$\gamma$ by a factor of $\sim 3$, the H$\beta$ trough looks only a bit deeper than the H$\gamma$ trough. This indicates that the H$\beta$ absorption line is saturated.
However, there is residual flux under the H$\beta$ trough, suggesting that the absorber partially covers its background source. The residual flux is greater than the predicted BEL flux of H$\beta$ (one can see this from Figure \ref{SED_fig} panel(a)), indicating the existence of remnant flux of the accretion disk, and therefore suggesting that the absorber covers the accretion disk partially. Considering that the size of the BELR is typically orders of magnitudes larger than that of the accretion disk, it is natural to suppose that the absorber barely covers the BELR.

Except for the redshifted absorption line system consisting of \ion{H}{1}*, \ion{He}{1}* and \ion{Fe}{2}*, we also found blueshifted \ion{Ca}{2} H\&K and \ion{Na}{1} D doublets in the optical spectrum. The \ion{Ca}{2} H\&K doublet, separated from each other by $2653~\mathrm{km~s}^{-1}$, are isolated. The equivalent widths (EWs) of the doublet can be measure directly as $1.43\pm 0.22~\mathrm{\AA}$ for \ion{Ca}{2} H and $3.42\pm 0.25~\mathrm{\AA}$ for \ion{Ca}{2} K, respectively. Considering the uncertainties, the ratio of the EWs is consistent with the theoretical ratio of the transition strength of 1:2, which suggests that these lines are unsaturated with a full coverage on its background source.
The \ion{Na}{1} D doublets are blended, so we measured their combined EW to be 3.9 \AA.
The origin of these blueshifted absorptions is evidently different from the redshifted system. Since neither was there any \ion{H}{1}*, \ion{He}{1}* or \ion{Fe}{2}* absorption identified in the blueshifted system, nor did we find any signs of \ion{Ca}{2} H\&K and \ion{Na}{1}D absorptions in the redshifted system, their physical conditions should also be different: the blueshifted absorber is cooler, and thus in lower ionization state.

\subsection{Measuring the Redshifted Absorption Lines}\label{Measure_Redshifted}

Since we supposed that the `blue' SED component comes from the scattering along the lines of sight different from that toward the central engine, it could have different absorptions with the redshifted absorption line system that we concerned about, which would complicate the measurement of the latter. Fortunately, the flux of the `blue' component drops quickly as the wavelength increases longward rest-frame $3000~\mathrm{\AA}$, and therefore its contribution to the fluxes around the redshifted and blueshifted absorption lines is trivial. In these sections we treat the quasar as a normal reddened quasar and apply the following measurements.

To measure the absorption lines, normalizing the spectra using the unabsorbed background fluxes is essential. The `pair-matching' method has been demonstrated as a practical method to provide satisfying guesses for unabsorbed background fluxes (Zhang et al. 2014, Liu et al. 2015, Shi et al. 2016b). Selected from a library of non-BAL quasars, one by one the individual BOSS spectrum was fitted to the spectral features of J1228+1005 surrounding the absorption lines, during which the absorption lines themselves were masked and the SMC-type extinction law was applied to account for the reddening. If the reduced $\chi^2<1.5$, we considered the fit acceptable, and the non-BAL quasar was regarded to have similar continuum and emission line characters with J1228+1005. The mean spectrum of these accepted non-BAL quasars' spectra were used as the unabsorbed template for J1228+1005, and the variance were used as the template's uncertainty (see Figure \ref{SED_fig} panel (a)).

Given the assumption that the BELR is not obscured by the redshifted absorber, the normalized absorption spectrum was obtained as follows:
\begin{equation}
F_\mathrm{norm} = \frac{ F_\mathrm{obs} - F_\mathrm{EL} }{ F_\mathrm{template} - F_\mathrm{EL} }
\end{equation}
where $F_\mathrm{obs}$ is the observed spectrum, $F_\mathrm{template}$ is the non-absorption template, $F_\mathrm{EL}$ is the total flux of the emission lines, and $F_\mathrm{norm}$ is the derived normalized spectrum.
Under the assumption of partial coverage, the normalized spectrum is related with the optical depth $\tau(\lambda)$ and the covering factor $C_{\mathrm{f}}$:
\begin{equation}
F_{\mathrm{norm}}(\lambda)=1-C_{\mathrm{f}}+C_{\mathrm{f}}\exp{(-\tau(\lambda))},\label{eq1} , \\
\end{equation}

\begin{figure}
\includegraphics[width=0.5\columnwidth]{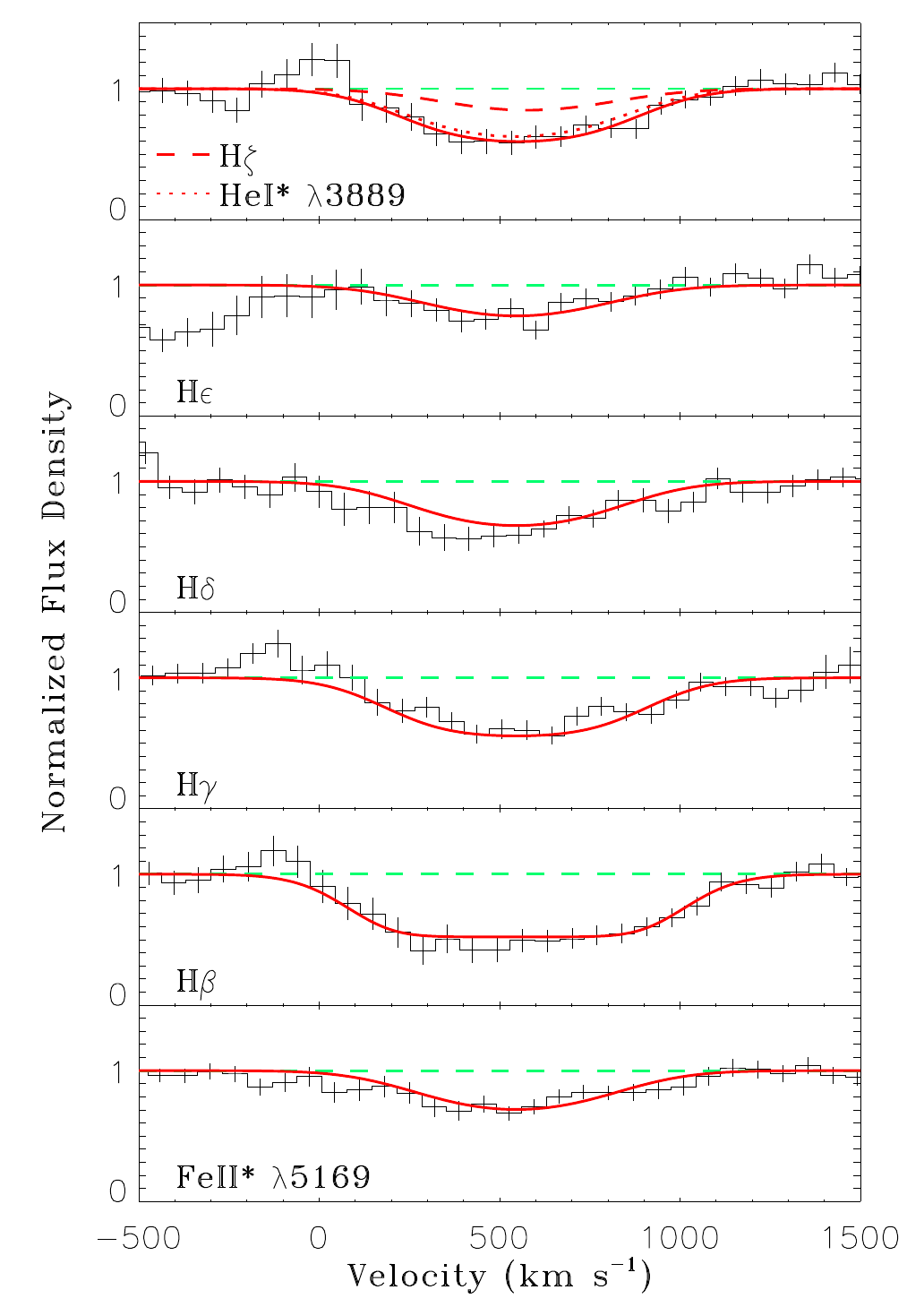}
\caption{The Gaussian profile fitting for the absorption lines in the redshifted absorption line system as functions of velocity shift with respect to the rest frame of J1228+1005. The black line is the normalized observational flux, the vertical bars show the normalized $1\sigma$ error, and the solid red lines are profiles of the fitting model. In the top panel, the dashed red line specifies the contribution of H$\zeta$, while the dotted red line specifies the contribution of \ion{He}{1}* $\lambda 3889$.\label{MeasRedAbs}}
\end{figure}

The mini-BAL troughs look symmetric and smooth, and thus a single Gaussian function can fit the velocity distribution of the optical depths well. The parameters can be expressed in terms of velocity: $v_{\mathrm{cen}}$ for the shift of the line center, and $v_{\mathrm{FWHM}}$ for the full-width-at-half-maximum (FWHM) of the Gaussian profile. Except for the H$\zeta$ which is blended with \ion{He}{1}* $\lambda 3889$, other Balmer absorption troughs were fitted simultaneously, using the same $v_{\mathrm{cen}}$ and $v_{\mathrm{FWHM}}$. The best fitting results were $v_{\mathrm{cen}}=546\pm 12~\mathrm{km~s}^{-1}$, $v_{\mathrm{FWHM}}=510\pm 22~\mathrm{km~s}^{-1}$, $C_{\mathrm{f}}=0.48\pm 0.03$, and the ionic column density of H$^0$ on the $n=2$ shell is $N(\mathrm{H}^0_{n=2})=2.74\pm 0.39\times 10^{15}~\mathrm{cm}^{-2}$ (see Figure \ref{MeasRedAbs}).

Given the column density of $\mathrm{H}^0_{n=2}$, the strength of H$\zeta$ would be known, and therefore its contribution to the blended trough at around $3889~\mathrm{\AA}$ can be accurately subtracted with the left absorption be attributed to \ion{He}{1}* $\lambda 3889$. Assuming all absorption lines of the redshifted absorption line system have the same profile, we used the Gaussian function with previous $v_{\mathrm{cen}}$ and $v_{\mathrm{FWHM}}$ to fit the residual \ion{He}{1}* $\lambda 3889$ and the weak \ion{Fe}{2}* absorptions: \ion{Fe}{2}* $\lambda 4523$, \ion{Fe}{2}* $\lambda 4550$, \ion{Fe}{2}* $\lambda 4585$, and \ion{Fe}{2}* $\lambda,\lambda 4925,5019,5170$. However, the absorptions of \ion{Fe}{2}* $\lambda 4523$, \ion{Fe}{2}* $\lambda 4550$, and \ion{Fe}{2}* $\lambda 4585$ are so weak that their normalized profiles are highly subjected to the uncertainty of the unabsorbed template. Therefore, we could only use the column density of Fe$^+$ on level a$^6$S$_{5/2}$ from \ion{Fe}{2}* $\lambda 5170$ in the following analysis. The results were $N(\mathrm{He}^0~(2^3\mathrm{S}))=1.17\pm 0.27\times 10^{15}~\mathrm{cm}^{-2}$ and $N(\mathrm{Fe}^+~(\mathrm{a}^6\mathrm{S}_{5/2}))=1.69\pm 0.25\times 10^{15}~\mathrm{cm}^{-2}$.

\subsection{Modelling the Redshifted Absorber}\label{Model_Redshifted}

The mere presence of the \ion{He}{1}* absorption in the redshifted system indicates that the absorber is illuminated by strong ionizing radiation, because the metastable level is populated through the recombination of He$^+$ ions. Therefore the redshifted absorber should be closer to the central engine than the blueshifted absorber, and be directly exposed to the radiation of the central engine. We used the photo-ionization simulation code CLOUDY (version 17.01, Ferland et al. 2017) to constrain its physical conditions. A Slab-shaped dust-free model for gas with homogeneous density and solar abundance was employed to reproduce the measured ionic column densities. The incident SED applied was a combination of a UV bump described as $\nu^{\alpha_{\mathrm{UV}}} exp(-h\nu/kT_{\mathrm{BB}}) exp(-kT_{\mathrm{IR}}/h\nu)$ and a power-law $a\nu^{\alpha_{\mathrm{X}}}$, which was extracted and combined from multi-band observations, and is considered typical for quasars (Ferland et al. 2017). The UV bump was parameterized by a UV power-law index $\alpha_{\mathrm{UV}}=-0.5$, and exponentially cut off with temperature $T_{\mathrm{BB}}=1.5\times 10^5~\mathrm{K}$ at the high-energy end and $T_{\mathrm{IR}}=1580~\mathrm{K}$ at the low-energy end. The power-law component had an index $\alpha_{\mathrm{X}}=-2$ beyond $100~\mathrm{keV}$ and $-1$ between $1.36~\mathrm{eV}$ and $100~\mathrm{keV}$. The overall flux ratio of the X-ray to the optical was $\alpha_{\mathrm{OX}}=-1.4$.

\begin{figure}
\includegraphics[width=\textwidth]{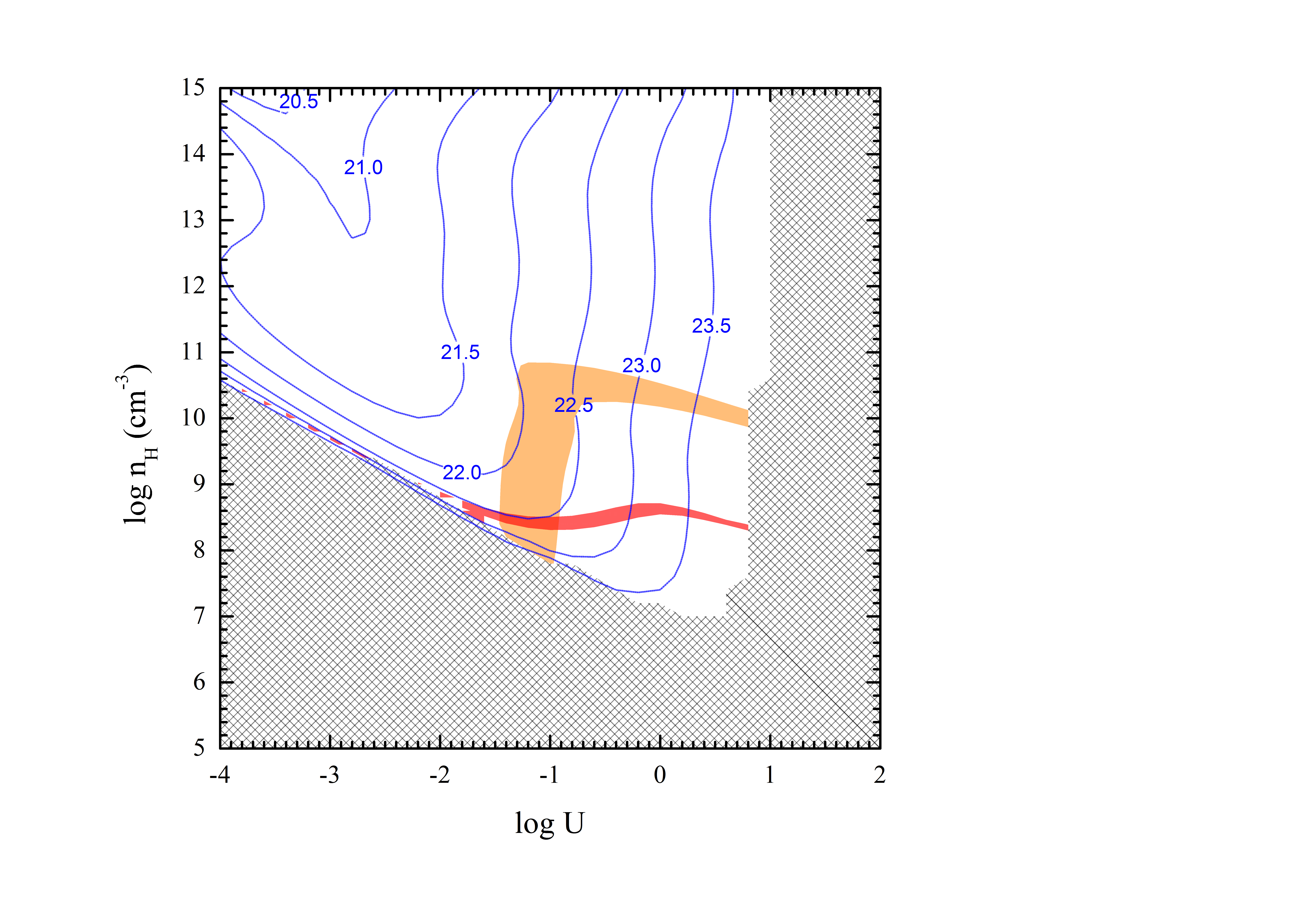}
\caption{The photo-ionization models evaluated using CLOUDY for given ionization parameters ($U$) and densities ($n_{\mathrm{H}}$), to constrain the physical properties for the redshifted absorption line system in J1228+1005. The blue contours present the total hydrogen column densities ($N_{\mathrm{H}}$) of H$^0_{n=2}$-selected models, which are defined as the preduced $N(\mathrm{H}^0_{n=2})$ equals to the measured value. The gridline shadowed area covers the $U$-$n_{\mathrm{H}}$ values for which no H$^0_{n=2}$-selected model can be found, since the measured $N(\mathrm{H}^0_{n=2})$ cannot be achieved whatever $N_{\mathrm{H}}$ is. The orange area presents the H$^0_{n=2}$-selected models for which the produced $N(\mathrm{He}^0~(2^3\mathrm{S}))$ values are consistent with the measured value within $1\sigma$ error, while the red area presents the H$^0_{n=2}$-selected models for which the predicted $N(\mathrm{Fe}^+~(\mathrm{a}^6\mathrm{S}_{5/2}))$ values are consistent with the measured value within $1\sigma$ error. Therefore, for models in the intersection, where $\log U=-1.2\pm 0.3$, $\log n_{\mathrm{H}}~(\mathrm{cm}^{-3})=8.4^{+0.2}_{-0.1}$, and $\log N_{\mathrm{H}}~(\mathrm{cm}^{-2})=22.55^{+0.10}_{-0.07}$, all the measured ionic column densities can be produced simultaneously.\label{PHImodel}}
\end{figure}

The most essential parameters to describe the photo-ionized gas are the photo-ionization parameter $U$, the total hydrogen density $n_{\mathrm{H}}$, and the total hydrogen column density $N_{\mathrm{H}}$. Given $U$ and $n_{\mathrm{H}}$, $N_{\mathrm{H}}$ can be adjusted to make the simulation produce the measured ionic column density $N_{\mathrm{col}}(\mathrm{H}^0_{n=2})$. So we built up a series of simulation models in which the value of $N_{\mathrm{H}}$ varys as a function of $U$ and $n_{\mathrm{H}}$ on the 2D $U$-$n_{\mathrm{H}}$ parameter plane. The ionic column densities of the detected He$^0_{2^3\mathrm{S}}$ and Fe$^+_{\mathrm{a}^6 \mathrm{S}_{5/2}}$ were also reproduced by these simulation models as functions of $U$ and $n_{\mathrm{H}}$. In Figure \ref{PHImodel}, we present the results of the simulations. The orange shaded area shows the region where the produced $N_{\mathrm{col}}(\mathrm{He}^0_{2^3\mathrm{S}})$ agrees with the measurement within 1-$\sigma$ uncertainty, and the red shaded area shows the region where the produced $N_{\mathrm{col}}(\mathrm{Fe}^+_{\mathrm{a}^6 \mathrm{S}_{5/2}})$ agrees with the measurement within 1-$\sigma$ uncertainty. The blue contours present the produced $N_{\mathrm{H}}$. At around $\log n_{\mathrm{H}}~(\mathrm{cm}^{-3})=8.4$ and $\log U=-1.2$, the orange area overlaps with the red area, which means the model here can simultaneous produce the measured column densities of H$^0_{n=2}$, He$^0_{2^3\mathrm{S}}$, and Fe$^+_{\mathrm{a}^6 \mathrm{S}_{5/2}}$. Therefore, the parameters of the best fit photo-ionization model were $\log U=-1.2\pm 0.3$, $\log n_{\mathrm{H}}~(\mathrm{cm}^{-3})=8.4^{+0.2}_{-0.1}$, and $\log N_{\mathrm{H}}~(\mathrm{cm}^{-2})=22.55^{+0.10}_{-0.07}$.

\begin{figure}
\includegraphics[width=\columnwidth]{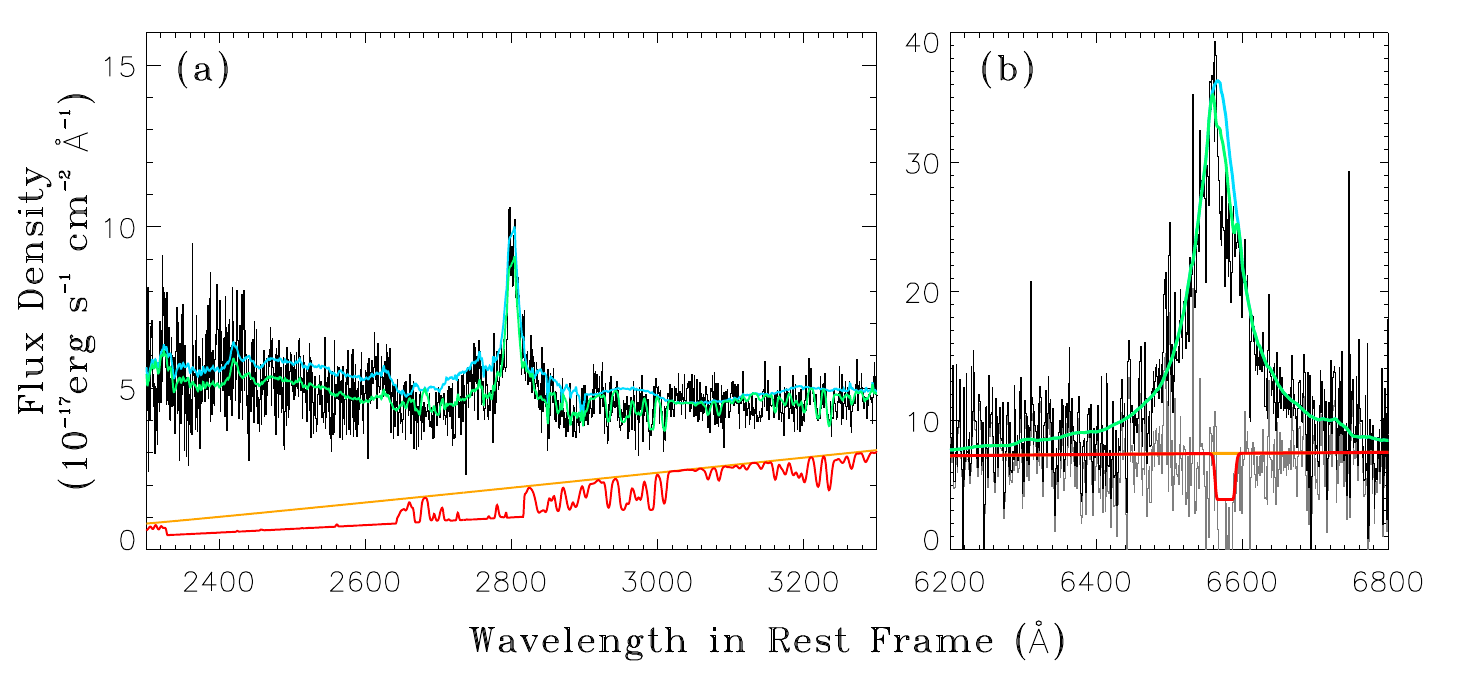}
\caption{Panel (a): The UV \ion{Fe}{2} and \ion{Mg}{2} absorptions predicted by the best-fit  photo-ionization model, compared with the SDSS spectrum. The dashed red line presents the unabsorbed continuum flux of the reddened component, while the solid red line presents the corresponding absorbed flux predicted by the best model. The dashed green line is the unabsorbed template for the overall spectrum, for which the corresponding model prediction (the solid green line) matches the observation. Panel (b): The TripleSpec NIR spectrum around H$\alpha$ emission. The dashed green line shows the best fit employing the SDSS quasar composite (see Figure \ref{SED_fig} panel (b)), and the predicted H$\alpha$ absorption is included in the solid green line. The observed flux is lower than the model trough. If we remove the emission flux, the difference would be more evident. The grey line shows the continuum flux only, while the solid red line shows the corresponding model spectrum.\label{OtherAbs}}
\end{figure}

The existence of the optical absorption lines from excited Fe$^+$ levels indicates a thick layer of low-ionized and neutral zone in the redshifted absorber. The best-fit photo-ionization model predicts that strong UV \ion{Fe}{2} (from the ground and the excited levels) and \ion{Mg}{2} $\lambda\lambda 2796,2803$ absorptions should be detected. However, since at the rest wavelengths $\lambda_{\mathrm{rest}} < 3000~\mathrm{\AA}$ the scattered flux overwhelms the penetrating (and thus reddened) flux, the absorber can only intercepts a small fraction of the total flux at these wavelengths, and therefore the predicted absorptions will be unapparent. In Figure \ref{OtherAbs} panel (a) we present the predicted absorption profiles. Although for the reddened component alone the absorptions are quite strong, its effect on the overall spectrum is little and can hardly be identified at a glance. For $\lambda_{\mathrm{rest}}<2820~\mathrm{\AA}$, the \ion{Mg}{2} and UV \ion{Fe}{2} absorptions are mostly saturated and blended, forming featureless and continuous absorption bands which indiscriminately decrease the overall fluxes by $\sim 15\%$. For \ion{Fe}{2} lines at around $2840\sim 3020~\mathrm{\AA}$ and $3200~\mathrm{\AA}$, the predicted unsaturated absorptions are found to be consistent with the observational features, supporting our photo-ionization model and the partial obscuration assumption. At even shorter wavelengths, the contribution from reddened component should be negligible, and we do not expect any sign of redshifted absorption lines, in other words \ion{Al}{3} $\lambda\lambda 1855,1863$, \ion{C}{4} $\lambda\lambda 1548,1550$, or \ion{Si}{4} $\lambda\lambda 1394,1403$ would not be detectable even if the UV spectrum were available.

\subsection{Measuring the Blueshifted Absorption Lines}\label{Blueshifted}

\begin{figure}
\includegraphics[width=0.5\columnwidth]{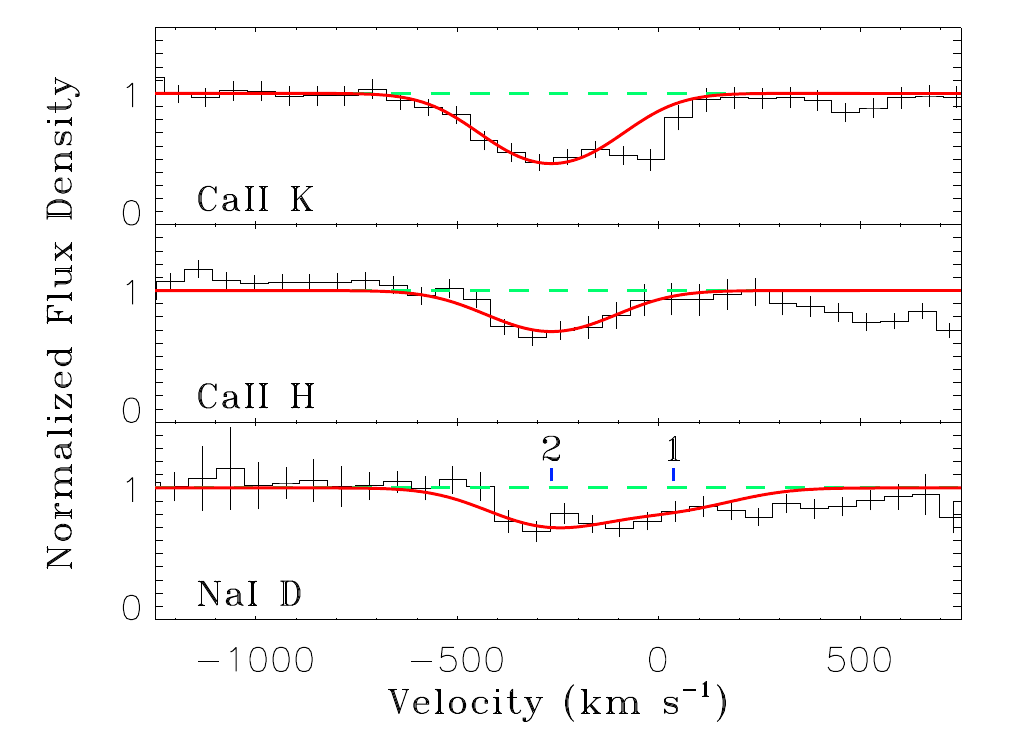}
\caption{The Gaussian profile fitting for the absorption lines in the blueshifted absorption line system as functions of velocity shift with respect to the rest frame of J1228+1005. The black line is the normalized observational flux, and the solid red lines are profiles of the fitting model. In the bottom, the line centers for both transitions in \ion{Na}{1}D are labeled.\label{MeasBlueAbs}}
\end{figure}

For the blueshifted absorption line system, the blue absorber was supposed to cover its background source fully, as indicated by the ratio of the EWs of \ion{Ca}{2} H\&K. Since there are no broad emission lines around the \ion{Ca}{2} and \ion{Na}{1} absorption lines, we can take the accretion disk only as the absorber's background source. Following Eq. \ref{eq1} and assuming that the optical depths of \ion{Ca}{2} H\&K have the same Gaussian profile, we fit the spectrum (see Figure \ref{MeasBlueAbs}), in which the the covering factor was initially set to be free. We obtained $C_{\mathrm{f}}=0.99\pm0.06$, which confirmed that the absorber fully covers its background source. Therefore we set $C_{\mathrm{f}}$ as 1 and fit the spectrum again. We got the absorber's blueshifted velocity of $227\pm 17~\mathrm{km~s}^{-1}$ with respect to the J1228+1005's rest frame, and the velocity dispersion was $v_{\mathrm{FWHM}}=343\pm 39~\mathrm{km~s}^{-1}$. The column density of \ion{Ca}{2} was $N_{\mathrm{col}}(\mathrm{Ca}^+_{\mathrm{ground}})=3.92\pm 0.40\times 10^{13}~\mathrm{cm}^{-2}$.
The \ion{Na}{1} D doublet, in which the lines' separation is only $304~\mathrm{km~s}^{-1}$, are blended. Using a Gaussian profile with the same $v_{\mathrm{cen}}$ and $v_{\mathrm{FWHM}}$ as \ion{Ca}{2} H\&K for each line and assuming $C_{\mathrm{f}}=1$, the blended absorptions can be reproduced, suggesting that \ion{Na}{1} D comes from the same absorber with \ion{Ca}{2} H\&K. The ionic column density was $N_{\mathrm{col}}(\mathrm{Na}^0_{\mathrm{ground}})=1.24\pm 0.16\times 10^{13}~\mathrm{cm}^{-2}$.

\section{Discussion}\label{Discussion}

\subsection{Estimating the Black Hole Mass and the Accretion Rate}

Using the parameters describing the broad band SED ($C_1, C_2, -\alpha_{\mathrm{scat}}$ and $\rm E(B-V)$, see \S\ref{SED}), we derived the intrinsic spectrum($ C_2 F_{composite}$) of J1228+1005.
Using this intrinsic spectrum we estimated the monochromatic luminosity at the rest-frame $5100~\mathrm{\AA}$ $L_{5100}=2.36\times 10^{45}~\mathrm{erg~s}^{-1}$.
Using the unabsorbed template for J1228+1005 (obtained by the pair-match method, see \S 4.1), we measured the FWHM of the unabsorbed H$\beta$ broad emission line $\mathrm{FWHM}(\mathrm{H}\beta)\sim 5.8\times 10^3~\mathrm{km~s}^{-1}$.
The mass of the central SMBH can be estimated according to the relation in Wang et al. (2009):
\begin{equation}
\log(\frac{M_{\mathrm{BH}}}{10^6~M_{\sun}})=(1.39\pm 0.14)+0.5\log (\frac{L_{5100}}{10^{44}~\mathrm{erg\ s}^{-1}})+(1.09\pm 0.23)\log (\frac{\mathrm{FWHM}(\mathrm{H}\beta)}{1000~\mathrm{km~s}^{-1}}).
\end{equation}
Using $L_{5100}$ and FWHM(H$\beta$) estimated above, we have $M_{\mathrm{HB}}\approx 8.1\times 10^8~M_{\sun}$.
The bolometric luminosity can be estimated as $L_{\mathrm{bol}}=(8.1\pm 0.4)\times L_{5100}=1.9\times 10^{46}\ \mathrm{erg\ s}^{-1}$ following Runnoe et al. (2012), and hence the Eddington ratio is about 0.19.
Assuming an accretion efficiency of 0.1, the mass accretion rate is then $\dot{M}_{\mathrm{BH}}\approx 3.3\ M_{\odot}\ \mathrm{yr}^{-1}$.

\subsection{Redshifted Absorber as Inflow}

The relation between the absorber's distance from the central engine $r_{\mathrm{abs}}$ and its physical conditions of $U$ and $n_{\mathrm{H}}$ is $\frac{L(<912)}{4\pi r_{\mathrm{abs}}^2}=Un_{\mathrm{H}}c\overline{E_{\mathrm{ph}}(<912)}$. $L(<912)$ is the ionizing luminosity of the continuum source, determined by the monochromatic luminosity at the rest-frame 912\AA  $L_{912}$ and the incident SED used in the photo-ionization simulation models. $\overline{E_{\mathrm{ph}}(<912)}$ is the average energy of all the hydrogen-ionizing photons, which can also be evaluated given the model SED. According to the flux of the unreddened intrinsic spectrum (Figure \ref{SED_fig} panel (b)) and $\log U=-1.2\pm 0.3$ and $\log n_{\mathrm{H}}(\mathrm{cm}^{-3})=8.4^{+0.2}_{-0.1}$ from the best photo-ionization model, the distance is $r_{\mathrm{abs}}=1.56^{+0.41}_{-0.58}~\mathrm{pc}$. The listed uncertainty only includes the uncertainties of $U$ and $n_{\mathrm{H}}$. The uncertainty of the AGN ionizing luminosity introduced by the model SED is more difficult to assess. For the simplest case, a change of 100\% in the luminosity can lead to a change of 41\% in the distance.

The distance above was calculated using cloudy simulations where the redshifted absorption-line inflow was assumed to be dust-free.
It is slightly larger than the evaporation radius for the dust grains, which could be roughly estimated as:
\begin{equation}
r_{\mathrm{evap}}=1.3 L^{1/2}_{\mathrm{UV},46}T^{-2.8}_{1500}~\mathrm{pc}\approx 0.98~\mathrm{pc},
\end{equation}
where $L_{\mathrm{UV},46}$ is the UV luminosity in unit of $10^{46}~\mathrm{erg~s}^{-1}$ estimated using the model SED's monochromatic luminosity $\lambda L_{\lambda}(1450)$ at rest-frame $1450~\mathrm{\AA}$, and $T_{1500}\sim 1$ is the grains' evaporation temperature in unit of $1500~\mathrm{K}$.
So the inflow gas may in principle contain dust.
Considering the fact that dust grains are easier to survive in low ionized or neutral gas, the dust on the line of sight ($E(B-V)=0.66$ mag) is more likely to be associated with the blueshifted absorber, which produces Na I and Ca II absorption lines.
Even if all this dust belongs to the redshifted absorber, the dust-to-gas ratio would be $1.9\times10^{-23}\ \mathrm{mag}\ \mathrm{cm}^{2}$, which is only about 70\% of that of the SMC (Golden et al. 2003).
Including such tenuous dust in the inflow would lead to two effects:
One effect is that dust and gas will compete for ionizing photons, resulting in a decrease in the degree of gas ionization.
The other effect is the depletion of metals into dust grains, resulting in a decrease in the iron abundance in the gas phase.
Assuming that all the line-of-sight dust belongs to the redshifted absorber, we conservatively estimated the greatest change to the distance brought by these effects using new Cloudy simulations.
In the simulations, the dust has a SMC type with grain size and composition set according to Weingartner \& Draine (2001), and the dust-to-gas ratio is set to be the upper limit estimated earlier.
We found that when $U$ is increased by 0.5 dex, the ionization structures of hydrogen and helium will be the same as that of dust-free models, and the column densities of H$^0_{n=2}$ and He$^0_{2^3\mathrm{S}}$ can be reproduced.
To compensate the shortage of Fe$^+_{\mathrm{a}^6 \mathrm{S}_{5/2}}$ caused by the decrease of Fe abundance in the gas phase, which is estimated to be $\sim$20\%, $n_H$ needs to be reduced by 0.1 dex and $N_H$ needs to be increased by 0.2 dex.
As a result, the distance will decrease by $\sim0.2$ dex at most.
These show that the dust will not have a significant impact on the distance of the redshifted absorber.
Despite of the large uncertainty in the distance of the absorber to the central engine, it is still on the same scale with the evaporation radius for the dust grains.

Reminding the scenario that inflows starting from the dusty torus feed the accretion disk (e.g., Krolik \& Begelman et al. 1988), this absorber could be a part of such accretion inflows that feeding the SMBH, similar to the absorbing gas with redshifted lines in other two  quasars SDSS J112526.12+002901.3 and J103516.20+142200.6 (Shi et al. 2016a; Zhou et al. 2019). Although it is not clear yet whether the inflows in quasars are continuous fluids or clumpy media, the mass of inflows can be estimated as:
\begin{equation}
M_{\mathrm{inflow}}=\mu m_{\mathrm{p}}N_{\mathrm{H}} 4\pi r_{\mathrm{abs}}^2 \Omega,
\end{equation}
where $\mu\approx 1.4$ is the mean atomic mass per proton, $m_{\mathrm{p}}$ is the mass of a proton, and $\Omega$ is the global covering factor of the inflows. Assuming that the global covering factor of the inflows is similar with that of the dusty torus and hence $\Omega\approx 0.6$, the mass of the inflows is about $3\times10^3\ M_\odot$.
Zhou et al. (2019) pointed out that for inflows feeding the accretion disk, the radiative pressure amounts to only a few percent of the gravitational force. If a cloud falls free from the inner torus, its infalling timescale would be $t_{\mathrm{ff}}\approx (r_{\mathrm{abs}}^3/GM_{\mathrm{BH}})^{1/2}=1\times 10^3~\mathrm{yr}$.
Therefore, the inflowing mass rate would be about $3 M_\odot\ \mathrm{yr}^{-1}$, which is close to the mass accretion rate $\dot{M}_{\mathrm{BH}}\approx 3.3\ M_{\odot}\ \mathrm{yr}^{-1}$ and would be sufficient to power the central engine.

\subsection{The Possible Variation of the Redshifted Absorption System}

The NIR TripleSpec spectrum obtained three years after the BOSS spectroscopic observation covers the H$\alpha$ emission peak in the object's rest frame. Following the same normalization process with the SDSS optical spectrum in fitting and measuring the redshifted absorption lines, we found that the residual flux in the normalized spectrum is around zero at the predicted wavelengths of the H$\alpha$ absorption. However, since $C_{\mathrm{f}}\approx 0.48$, the predicted absorption profile should be much shallower (see Figure \ref{OtherAbs} panel (b)). If the inconsistency were confirmed, the absorption line system would have experienced considerable variation during the five years. Unfortunately, we have no simultaneous optical spectrum to measure the optical lines in 2017, and the \ion{He}{1}* $\lambda 10830$ trough in the TripleSpec spectrum falls in the gap around $1.85~\mu\mathrm{m}$ due to telluric absorptions. No other observation features can be used to confirm the variation.

The possible variation could be explained in two pictures. It could be driven by a change in the photo-ionization state (Sun et al. 2017, He et al. 2017) or a transverse motion (Shi et al. 2016b). The absorber could be clumpy, in which cores of higher densities are surrounded by diffuse medium of lower densities (Hamann et al. 2001; 2019). During the SDSS observation in 2012, the diffuse medium was fully ionized. Low ionized and neutral gas could only survive in the cores, generating the observed redshifted absorption line system with $C_{\mathrm{f}}\approx 0.48$. If the central engine dimmed from then on (although we do not find such a trend in Catalina $V$-band photometry from 2005 to 2013), new low ionized and neutral gas would emerge in the diffuse medium and enlarge the obscured fraction of the background source, which would result in a larger $C_{\mathrm{f}}$ for absorption lines observationally. Also, the dynamic structure of the diffuse medium would be different from that of the cores (usually the velocity dispersion should be larger). Therefore, the H$\alpha$ absorption rising from the diffuse medium should have a trough wider than that from the cores. However, in the NIR spectrum the width of the H$\alpha$ absorption line is the same with those of the Balmer absorption lines in the SDSS spectrum. Hence the explanation that the variation is driven by photo-ionization change is not preferred.

The transverse motion of the absorber presents the other explanation for the variation in the absorption. When the cloud starts to move across the line of sight, at this stage it will gradually block the background source, with its covering fraction increases and the absorption troughs deepen, while the apparent optical depth ratio of H$\alpha$/H$\beta$ will remain as the photo-ionization state is unchanged. Since the dynamics inside this obscuring cloud changes little when $C_{\mathrm{f}}$ increases from 0.48 to 1, the widths of the absorption lines will also remain. This picture seems to be more consistent with the observation.

The two pictures might be distinguished by X-ray observations, because the transverse motion can cause a change in the X-ray absorption column density, while the effect of change in the photo-ionization state on the X-ray spectrum is different. However we could not make this because the quasar was not detected by ROSAT/PSPC or SWIFT/XRT observations that covered it.

\subsection{The Dusty Medium on the Line of Sight}

In \S\ref{SED} we found that the quasar is reddened with $E(B-V)=0.66$, indicating large amount of dust on the line of sight. Gas surrounding such dust is generally low ionized or even neutral, facilitating the absorptions of \ion{Ca}{2} and \ion{Na}{1}.
Poznanski et al. (2012)(hereafter P12) gives empirical relations between the color excess and the EWs of \ion{Na}{1}D absorption lines of the ISM in the Milky Way. Following the relations in P12, a reddening with $E(B-V)=0.66$ corresponds to a combined $\mathrm{EW}(D_1+D_2)$ of 1.5 \AA.
In J1228+1005 the blueshifted absorber may be associated with the dust on the line of sight, but the \ion{Na}{1}D absorption lines it produces have combined $\mathrm{EW}(D_1+D_2)$ of 3.9 \AA, much larger than that predicted by P12.
A possible reason for this disagreement lies in the large velocity dispersion of the blueshifted absorber. In \S 4.3 we obtained its velocity dispersion as $\sim$340 km s$^{-1}$. If we were to modify it to $\sim$10 km s$^{-1}$, which is the common velocity dispersion of the ISM in the Milky Way, the combined $\mathrm{EW}(D_1+D_2)$ would drop to 1.4\AA, in agreement with that predicted by P12. Therefore, it is reasonable to assume that the blueshifted absorber is associated with the dust on the line of sight.

Not only the velocity dispersion of the blueshifted absorber (FWHM$\sim$340 km s$^{-1}$) is larger than that of general ISM, but also the blueshifted velocity (220 km s$^{-1}$) is larger than that of the proper motion of bound ISM.
There are two possible explanations for these phenomena.
The first is that the blueshifted absorber may be part of the dusty torus. Its large velocity dispersion is caused by the high-pressure environment in the galactic center, and its blueshifted velocity is consistent with Kepler velocity at $\sim$10 pc for a SMBH with $8\times10^8$ $M_\odot$.
The second is that the blueshifted absorber may be a part of an outflow. Its blueshifted velocity and velocity dispersion are caused by the acceleration by shock or radiation pressure. Because the profiles of the \ion{Ca}{2} and \ion{Na}{1} absorption lines look symmetrical and can be fitted by a single Gaussian function, the first explanation is preferred.
Since the line of sight passes through an inflowing cloud, and a dusty cloud that might be associated with the torus, the quasar J1228+1005 is likely to be observed from a large inclination angle.

We found, via SED fitting in Section 3, a strong scattered light in J1228+1005 by detection of a significant UV excess,
and suggested that such a flux excess might be originated from quasar continuum scattered by dust grains.
The scattering mirror candidate could either be the surface of the dusty torus, or be the polar dust material locating inside the ionization cone, as suggested by infrared interferometric studies (e.g., Raban et al 2009; H\"{o}nig et al 2013; Tristram et al 2014).
The light scattered by the former is easier to be reddened by outer dust, especially in a system that may have a large inclination angle.
Thus, as the observed scattered light is relatively blue and seems not reddened, it is more likely to be from the polar dust.

\subsection{An Overall Picture of the Redshifted and Blueshifted Absorbing Media}

\begin{figure}
\includegraphics[width=\textwidth]{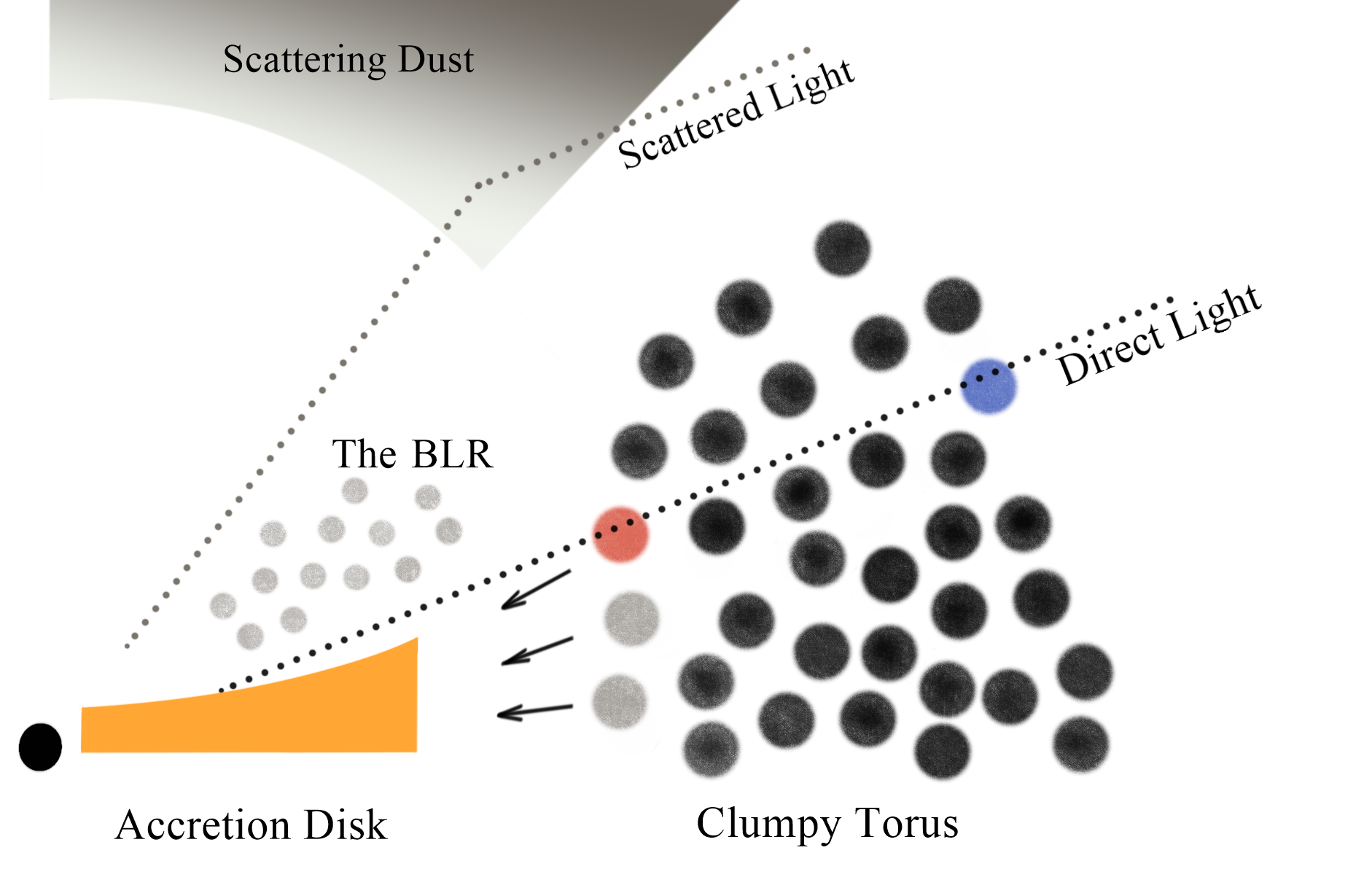}
\caption{A quadrant illustration of the geometry of the quasar nucleus and the absorbers. The direct light (dotted black line) penetrates the dusty torus, in which it is intercepted by the redshifted absorber (red circle) and skims over the blueshifted absorber (blue circle). The redshifted absorber is near the inner surface of the clumpy torus, where gaseous clouds (big grey circles) could be falling inward. The blueshifted absorber is far away from the central engine, and it could be one of the dusty clouds (big black circles) in the clumpy torus.
In addition, the quasar's light can also be seen through the scattering of the polar dust.
\label{Illustration}}
\end{figure}

We present a brief illustration of our understanding of the central engine and the absorbers in J1228+1005 in Figure \ref{Illustration}.
The reddening due to dusty medium is the major difference between J1228+1005 and the other two well-studied quasars hosting absorption-line inflow, i.e. J112526.12+002901.3 and J103516.20+142200.6.
Since the adjacent inflows are considered lying close to the mid-plane of the central engine, serious extinction due to the dusty torus should be expected for the lines of sight towards the inflow.
In the case of J1228+1005, dusty medium is indeed present on the line of sight, as revealed by its reddening and by the blueshifted absorption line system.
Thus, J1228+1005 might present a more typical example for quasars hosting absorption-line inflows.

Such an idea is consistent with the fact that redshifted absorption line system representing adjacent inflow is rare in observation. For a line of sight with large inclination angle with respect to the mid-plane of the central engine, the inflow would not intervene it, while for a line of sight close to the mid-plane which passes through both the inflow and the bulk of the torus, no radiation directly from the central engine could be observed in rest-frame UV and optical bands. Only when occasionally a line of sight passes through the hole of the dusty torus can the inflow be detected via redshifted atomic absorption lines.
According to the mainstream torus models (e.g., Nenkova et al. 2008), the dust in the torus is highly clumpy: most of the dust is in optically thick clouds (V-band optical depth $\tau_V\sim100$), and on a line of sight the number of such clouds $\mathrm{N}$ is on average 5 to 15.
Assuming that $\mathrm{N}$ is 10, the probability that there is no such cloud along a line of sight will be $\sim e^{-10}\sim5\times10^{-5}$.
Although the probability is small, in the SDSS/BOSS surveys which provide $\sim10^6$ sources one is possible to find such exceptional cases toward which the lines of sight pass through the holes in their dusty tori. In these cases, there may be diffused dusty medium or gas clouds with moderate amount of dust on the lines of sight, redden the sources' spectra and leave \ion{Ca}{2} and \ion{Na}{1} absorption lines like that in J1228+1005, instead of optically thick clouds totally obscuring the quasar.

The scattering by ambient medium complicates the situation. For high-$z$ quasars it is possible that what we observe in the rest-frame optical band are in fact scattered fluxes, while their central engines suffer serious extinction by the dusty medium on the lines of sight and could only be detected in the IR band. In this situation, a luminous quasar hosting redshifted absorption line system from massive inflows might be identified as a fainter none-BAL blue quasar through optical spectroscopic observation. This could also be one of the reasons for why redshifted absorption lines from inflows are so rarely observed in quasars. A spectroscopic survey in IR band can minimize such bias.

\section{Summary}\label{Summary}

The redshifted absorption line systems, especially those consisting of \ion{H}{1}* and \ion{He}{1}* lines, present powerful diagnostic to the inflows adjacent to the central engines of AGNs. Following our previous work, the redshifted \ion{H}{1}* and \ion{He}{1}* absorptions in J1228+1005 were identified tracing another inflow gaseous medium of high density and high column density at $\sim$1.6 pc away from the central black hole, and could also be originated from the inner torus. However, the line of sight is a bit different as it suffers considerable reddening due to the dusty medium which was suggested to be associated with a blueshifted low-ionization absorption line system. The properties of this blueshifted system are consistent with what we expect for the medium of the torus. If the dusty medium is really a part of the clumpy torus, the object will be remarkable for allowing the detection of the inflow with severe local extinction exists, which should be intrinsic for massive accretion inflows considering the theoretical guess on their geometry. The guess also expects that only in a narrow window for inclination the absorption inflow is accessible, which is consistent with the phenomenon that the J1228+1005 analogue is rare in observation.

\begin{acknowledgements}
This work is supported by the Strategic Priority Research Program of Chinese Academy of Sciences, Grant No. XDB34000000.
X.-H. S. is supported by Shanghai Natural Science Foundation (20ZR1463400).
This research uses data obtained from the MAST.
This research also uses data obtained through the Telescope Access Program (TAP).
Observations obtained with the Hale Telescope at Palomar Observatory were obtained as part of an agreement between the National Astronomical Observatories, Chinese Academy of Sciences, and the California Institute of Technology.
\end{acknowledgements}

\clearpage

\clearpage

\clearpage

\clearpage

\clearpage

\clearpage

\clearpage

\end{document}